# Identifying Depression on Twitter

Moin Nadeem[1], Mike Horn.[1], Glen Coppersmith[2], *Johns Hopkins University* and Dr. Sandip Sen[3], *PhD, University of Tulsa*

[1]Advanced Placement Research, Jenks High School, Jenks, OK 74037, USA
[2]Department of Computer Science, Johns Hopkins University, Baltimore, MD 21218, USA
[3]Department of Computer Science, University of Tulsa, Tulsa, OK 74104, USA

**Social media has recently emerged as a premier method to disseminate information online. Through these online networks, tens of millions of individuals communicate their thoughts, personal experiences, and social ideals. We therefore explore the potential of social media to predict, even prior to onset, Major Depressive Disorder (MDD) in online personas. We employ a crowdsourced method to compile a list of Twitter users who profess to being diagnosed with depression. Using up to a year of prior social media postings, we utilize a Bag of Words approach to quantify each tweet [1]. Lastly, we leverage several statistical classifiers to provide estimates to the risk of depression. Our work posits a new methodology for constructing our classifier by treating social as a text-classification problem, rather than a behavioral one on social media platforms. By using a corpus of 2.5M tweets, we achieved an 81% accuracy rate in classification, with a precision score of .86. We believe that this method may be helpful in developing tools that estimate the risk of an individual being depressed, can be employed by physicians, concerned individuals, and healthcare agencies to aid in diagnosis, even possibly enabling those suffering from depression to be more proactive about recovering from their mental health.**

## I. Introduction

Mental health is and continues to be a prominent plague for the civilized world. It is estimated that one in four American citizens suffers from a diagnosable mental disorder in any given year [1]. When combined with the 2015 US Census for Residents 18 and older, these statistics create a picture of 80 million suffering United States citizens [2]. One in three of these citizens who suffer from a mental illness may suffer from clinical depression, thus launching a wealth of studies to tackle this matter [3]. From this substantive field, we choose to focus on Major Depressive Disorder, commonly referred to as clinical depression.

Not only do nearly 300 million people worldwide suffer from clinical depression, but the probability for an individual to encounter a major depressive episode within a period of one year is 3 – 5% for males and 8 – 10% for females [4]. Yet, these effects of depression reach further than simply societal happiness: Depression takes a toll on U.S. businesses that amounts to over $70 billion annually lost in medical expenditures, productivity, and similar costs. An additional $23 billion in other costs may accrue on behalf of an individual, thus affecting workdays, diminishing work habits, and potentially inciting complications with concentration, memory, and decision-making behaviors.

This also stems further than simply the economic sphere, and often co-occurs with other illnesses and mental conditions. One in four cancer patients experience depression, one in three heart attack survivors undergo depression, and up to 75% of individuals diagnosed with an eating disorder will encounter the disease [5]. Major depressive disorder has also presented itself as the leading cause of disability worldwide among individuals five and older, and has been correlated with a higher risk for broken bones in women [6]. While these factors may be seemingly absurd on their own, they often coalesce into an undue burden on an ailing patient, thus vastly degrading the quality of life for an individual and their peers.

Any individual suffering through one or more of these mental illnesses will likely experience a snowball effect towards others, therefore exponentially increasing the likelihood of suicide [7]. Levine et al. determined that three people commit suicide for every two that are involved in a homicide, thus reinforcing our claim on the validity of the problem at hand [8]. Nevertheless, over two-thirds of 30,000 suicides reported in the past year were due to depression [9]. Mann et al. determined that untreated depression is the number one risk for suicides among youth, and suicide itself is the third leading cause of death among children [10]. Clearly, depression has the potential to manifest itself within a cornucopia of other social issues, and therefore becomes a problem of high priority for our society to solve.

However, current methods to identify, support, and treat clinical depression have been considered inefficient. Only 87% of the governments in the world provide some form of primary care to combat mental illnesses, and 30% of world governments provide no institution at all for mental outreach [11]. The bulk of the complications with a mental health diagnosis lie within the fact that no laboratory test has been created; the diagnosis is simply derived from a patient's self-reported experiences, behavior questionnaires, surveys, and a single mental health status examination.

These examinations for estimating the degree of depression within individuals are typically administered in the form of questionnaires, which vastly vary in form and length; the most popular of these are the Center of Epidemiologic Studies Depression Scale (CES-D) [12], Beck's Depression Scale (BDI) [13], and Zung's Self-Rating Depression Scale (SDS) [14]. Results on these examinations are commonly determined from the patient themselves, or a third-party observation, but never from empirical data. Thus, these questionnaires often lend themselves flaws though subjective human testing, and may be easily manipulated to achieve a pre-determined prognosis. These results often coincide with effort to gain anti-depressants, or otherwise mask one's depression from a friend or family member. Furthermore, these questionnaires are often costly, and further the economic burden of receiving treatment for depression significantly.

Yet, before even examining flaws with the current methods of treating depression, one must simply be identified with the illness: the World Health Organization reports that the vast majority of depressed individuals never seek out treatment [15]. This is particularly troubling for the younger generation, which commonly will resort to blame and stifled self-esteem before seeking help. Even during visits with a primary health care physician, depression often goes unrecognized, and therefore undiagnosed [16].

Yet, when Major Depressive Disorder (MDD) is properly identified, contained, and treated, it may have far-reaching impacts upon society. Up to 80% of those treated for depression showed an improvement in their symptoms within four to six weeks [10], thus bettering their lives, productivity, and boosting the economy. A study funded by the National Institute of Mental Health developed a test to determine the effectiveness of depression treatment. Known as the Sequenced Treatment Alternatives to Relieve Depression (STAR*D): it reported depression remission rates of over 65 percent after six months of treatment [17]. Therefore, it has become patently obvious that our largest contribution to combating clinical depression in the United States would lie within improving techniques to identify Major Depressive Disorder, rather than in its treatment methods.

**We therefore explore, develop, and test an algorithm to identify Major Depressive Disorder (MDD) via social media.** People increasingly utilize social media platforms to share their innermost thoughts, desires, and voice their opinion on social matters. Postings on these sites are made in a naturalistic manner, and therefore provides a solution to the manipulation which self-reported depression questionnaires often encounter. We have concluded that social media provides a means to capture an individuals present state of mind, and is even effective at representing feelings of worthlessness, guilt, helplessness, and the levels of self-hatred that would often characterize clinical depression. We pursue the hypothesis that social media, through word vectorization, may be utilized to construct statistical models to detect and even predict Major Depressive Disorder, and possibly even compliment and extend traditional approaches to depression diagnosis.

Our main contributions to this paper are as follows:

1) We utilize a dataset created by Coppersmith *et al.* for the Computational Linguistics and Clinical Psychology (CLPsych) 2015 Shared Task. The data was collected from Twitter users who stated a diagnosis of depression, and then was normalized across a standard demographic distribution. Each user in this dataset is anonymized for privacy purposes.
2) We sift through the provided models to refine several statistical measures and used them to quantify an individual's social media behavior across the maximum of three thousand tweets, as recommended by Tsugawa et al [18].
3) We compare the behaviors of the depressed user class that of the standard, and utilize a Bag of Words approach to tokenize the models provided by the CLPsych Shared Task. We process the tokenization through a Vectorizer to quantify the features.
4) We leverage the signals derived from the Bag of Words approach to develop, and contrast several MDD classifiers, and provide a statistical analysis to evaluate the results of each one. Our best model demonstrated promise in predicting the mental health condition of a user with an accuracy of 82% and a precision score of .86.

This research is novel in its use as foundational Text Classification system, while many other common projects provide an analysis as to whether or not a user is depressed, we focus our reach onto whether their tweets are depressive in nature or not on a document-level basis. We believe this research could further the underlying infrastructure for new mechanisms which may identify depression and related variables, and may even frame directions which could guide valuable interventions for a user. Ultimately, we desire for this research to be built upon with increased features, therefore leveraging the power of statistical models to save lives.

II. LITERATURE REVIEW

Rich bodies of work on depression have been performed within the psychiatry, psychology, medicine, and sociolinguistic fields to identify and correlate Major Depressive Disorder and its symptoms. In the areas of medicine and psychology, several questionnaire-based measures rating depression have been proposed. CES-D [12], BDI [13], and SDS [14] estimate the severity of depression in individuals from a variety of self-reported answers to questionnaires ranging from 17 questions to 20 questions. Yet, few approaches utilize objective information to determine their prognosis.

Redei et al. has discovered biological markers for early-onset Major Depressive Disorder, which could increase specify in the diagnosis for clinical depression. Their analysis of 26 candidate's blood transcriptomic markers in a sample of 15 – 19 year-old subjects resulted in a correct diagnosis for 11 out of 14 candidates who suffered from depression; another panel was able to distinguish between MDD or comorbid anxiety for 18 individuals [19]. Redei *et al.* is notable for being the first significant approach towards identifying depression from a medical perspective, although numerous exist from a data science perspective [19].

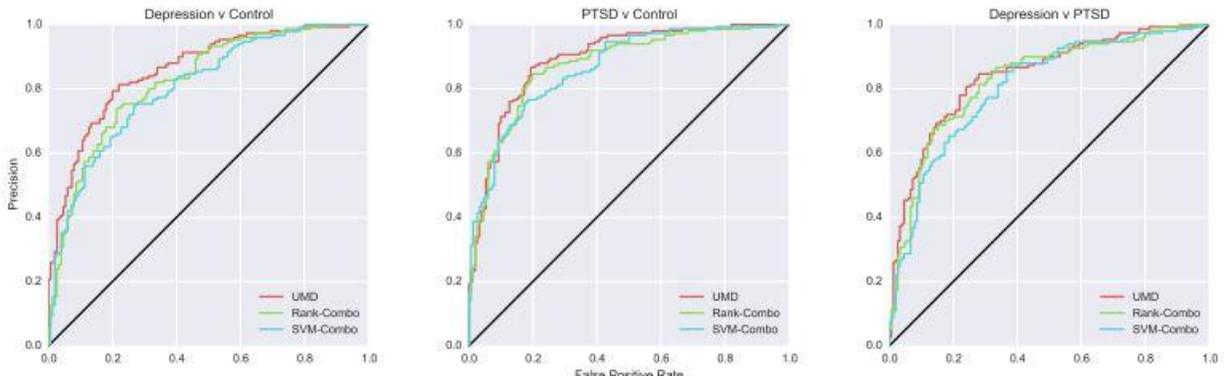

**Figure 1 shows the ROC curves for each particular model for each combination of classes. In particular, University of Maryland's approach consistently outperformed competitors.**

Approaches that utilize objective information, such as log data about an individual's activities to predict depression have been studied recently. Resnik *et al*. has formulated a method for identifying depression in individuals through analyzing textual data written by these individuals. They obtained topics from the essays written by college students by applying latent Dirichlet allocation (LDA), a popular topic-extraction model within Machine Learning [20]. Through using these discovered topics from a statistical model, they were able to estimate depression and neuroticism in college students with an *r* value of .45, thus discovering a slight correlation between neuroticism, depression, and academic works by college attendees. Resnik *et al.* becomes relevant for their novel use of topic modeling; otherwise these academic works are often a poor dataset to derive diagnoses from [20].

If not academic papers, researchers have discovered a deep correlation between the troves of data available under social media profiles and depression diagnoses. In 2011, Moreno *et al*. selected profiles from social media mogul Facebook, and evaluated personally written bodies of text, henceforth referred to as a 'status update' [21]. Through a set of criteria standardized by the American Psychiatric Association known as Diagnostic Criteria for Major Depressive Disorder and Depressive Episodes (DSM), they were able to determine that 25% of 200 selected profiles displayed signs of depression, and 2.5% met criteria for a Major Depressive Episode. Park *et al*. discovered the same method may be utilized for Twitter through analyzing Twitter users with and without depression and their online activities. Using a simple regressive analysis, Tsugawa *et al*. discovered that frequencies of word usage are useful as features towards identifying depression on Twitter, and therefore is notable for furthering the search of which features to utilize in order to estimate the severity of depression [22].

The most prominent of these studies have been conducted by de Choudhury *et al*., who have pioneered several novel aspects into the statistical models they develop. Most prominently, the specificity of their research has notably differentiated from many others in the field, as they focused on providing an estimate as to risk of depression through the user's behavior, rather than their status updates [23]. In their study, they developed the usage of features such as emotion, egonetwork (ie. monitoring the social activities of one with their close friends), linguistic styles, depressive language, and demographics to feed into their statistical model [23]. De Choudhury *et al*. discovered that the onset of depression through social media may have been able to be characterized through a decrease in social activity, raised negative effect, a highly clustered egonetwork (ie. highly clustered social groups, as opposed to an open-graph model), heightened relational and medicinal concerns, and a greater expression of religious involvement. Fed into a Support Vector Machine, a dimension-reduced coalition of these features offered a 72% accuracy rate, thus toppling anything pre-existing research within the field [23].

Lastly, Coppersmith *et al*. developed a Shared Task for the Computational Linguistic and Clinical Psychology (CLPsych) conference. Through this shared task, Coppersmith *et al*. distributed a standardized dataset of depressed, Post-Traumatic Stress Disorder (PTSD), and control users to all competitors in order to normalize fundamental computational technologies which often were at play [24]. This shared task was distributed to several participants, including the University of Maryland [25], the University of Pennsylvania's World Well-Being Project [26], the University of Minnesota-Duluth, and MIQ [24] (a small team composed of Microsoft, IHMC, and Qntfy). As we work with the dataset from Coppersmith *et al* [24], it is imperative to summarize each competitor's contributions below.

1) UMD utilized a supervised topic model approach to discover groupings of words that provided maximal impact to differentiate between the three provided classes for each user. Furthermore, rather than treat each tweet as its own document, or treat each user as one collective document, they chose to sensibly concatenate all tweets from a given week as a single document [20].
2) The WWBP team utilized straightforward regression models with a wide variety of features, including inferring topics automatically, and binary unigram vectors (ie. "did this user ever tweet this word?"). These topic models provided varying interpretations on which groups of words belonged together, thus providing insight as to which approach best expresses mental health-related signals [26].
3) The team from Duluth took a powerful approach to this by decoupling the power of an open-vocabulary approach

to simple, raw language features. Quite importantly, this open vocabulary approach might have been simplistic in nature but achieved an average precision in the range of .70 - .76, while complex machine learning or complex weighting schemes performed just as well [27].

4) The Microsoft-IHMC-Qntfy joint team utilized a character language model (CLMs) to determine how likely a given sequence of characters is to be generated by each classification class, and provided a score for each string. The beauty of this approach lied within scoring extremely short text, capturing information for creative spellings, abbreviations, and other textual phenomena which derives from Twitter's unique 140-character limit [24].

Our study builds upon prior mentioned work and contributes towards enhancing lexical methods for text classification. With our present work we: (1) further explore the capability for individual social media status updates to be utilized as a feature in determining or furthering a diagnosis of depression or not; (2) examine, compare, and analyze the effectiveness of several supervised statistical models to predict text classification; and (3) demonstrate that we may use these features to further the identification of depressive disorders in a cohort of individuals who may otherwise have slipped under the radar.

### III. AIMS

This study aims to establish the feasibility of consistently detecting, identifying, and pursing the diagnosis of individuals Twitter posts, henceforth referred to as 'tweets', Using solely these tweets, we aim to design and implement an automated computational classifier which may be able to parallel the performance and precision of a concerned human individual. The feasibility of this automated predictions will be cross-validated and critiqued through standard Precision, Recall, and F1 scores, as well as Recipient Operating Classification curves. The definition of these metrics are as follows:

#### A. Precision

$$\text{precision} = \frac{|\{\text{relevant documents}\} \cap \{\text{retrieved documents}\}|}{|\{\text{retrieved documents}\}|}$$

Precision is the fraction of retrieved documents that are relevant to the query. In our circumstances, it answers the question: *"How many of the users we identified as depressed are actually depressed?"*

#### B. Recall

$$\text{recall} = \frac{|\{\text{relevant documents}\} \cap \{\text{retrieved documents}\}|}{|\{\text{relevant documents}\}|}$$

Recall is the probability that a relevant document is retrieved by the query. Within our situation, it answers the question *"Out of all of the depressed users, how many did we properly detect?"*

#### C. F1 Score (F-Measure)

$$F_1 = 2 \cdot \frac{\text{precision} \cdot \text{recall}}{\text{precision} + \text{recall}}$$

An F1 score is the harmonic mean of Precision and Recall; it therefore is commonly utilized as a classification evaluation metric due to weighing each metric evenly.

### IV. METHOD

#### A. Data

In this study, we gathered information from the Shared Task organizers of the CLPsych 2015 conference. This dataset was developed from an amalgamation of users with public Twitter accounts who posted a status update in the form of a statement of diagnosis, such as "I was diagnosed with X today", where X would represent either depression or PTSD [24].

For each user, up to 3000 of their most recent public tweets were included in the dataset, and each user was isolated from the others. It should be noted that this 3000 tweet limit derives from Twitter's archival polices [22], and that most tweets concentrated long after a two-month timespan may possibly lower the effectiveness of a classifier, as shown by Tsugawa *et al.* [18].

Before releasing the dataset to participants, Coppersmith *et al.* matched age and gender to the demographics of the population, ultimately designing a dataset which consisted of 574 individuals (~63% of the dataset) with no mental health condition, and 326 users (~36% of the dataset) with a mental health condition of depression. For the purposes of our research, this resulted in 1,253,594 documents (tweets) as control variables, and 742,560 documents with a mental health condition of depressed.

Lastly, each user, and the users they interacted with, had been anonymized the dataset to ensure their privacy would be protected. In addition, Shared Task participants were required to sign a privacy agreement, institute security measures on the data, and obtain the approval of an ethics review board in order to secure the dataset. Data had been distributed in compliance with Twitter company policy and terms of service.

#### B. Features

We next present a set of attributes which can be used to characterize the behavioral and linguistic differences of the two classes – one of which consists of tweets which exhibits behavior reflective of clinical depression. Note that these measures are on a document-level i.e. rather than treating each user as a single unit, we examine each tweet as its own isolated document. Therefore, our research assumes a more granular scope, and becomes unique in nature.

We utilize a Bag of Words approach, which utilizes word occurrence frequencies to quantify the content of a tweet, i.e. putting all words within a bag and measuring how commonly each word appeared. Tsugawa *et al.* demonstrated that a Bag of Words approach can be quite useful for identifying depression, as he obtained a maximum *r* correlation

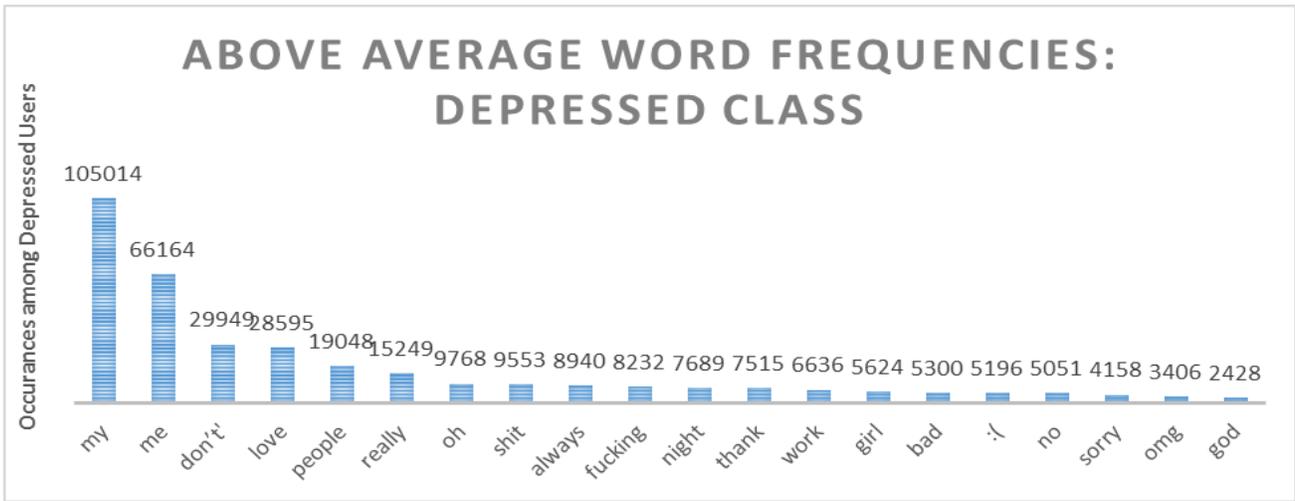

**Figure 2 demonstrates abnormally high word occurrences within the depressed class.**

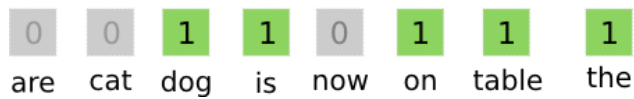

**Figure 3. A visual representation of the Bag of Words approach: we attempt to quantify depression through an analysis of word frequencies.**

coefficient of 0.43 among common words [18]. Fig 3 depicts some words which were more profound within the depressed class than the control, as well as their frequencies.

Many within this field decide to normalize term frequencies based off of document length, however, we determined no quantitative advantage during preliminary testing as the document length was consistent between tweets. By feeding these into our statistical model, our goal is to quantify depression, and ultimately an estimate as to the likelihood of depression within an individual.

*C. Classifiers*

We employ four different types of binary classifiers in order to estimate the likelihood of depression within users. For each classifier, we utilize Scikit-Learn from Pedregosa et al. to implement the learning algorithm [32]. We chose to evaluate Linear, Non-Linear, and Tree-based approaches in order to shallowly explore foundational learning models against our dataset. Ultimately, we decided upon Decision Trees, a Linear Support Vector Classifier, a Logistic Regressive approach, as well as a Naïve Bayes algorithm. In this section, we attempt to explain how these algorithms work, as well as our implementation of them.

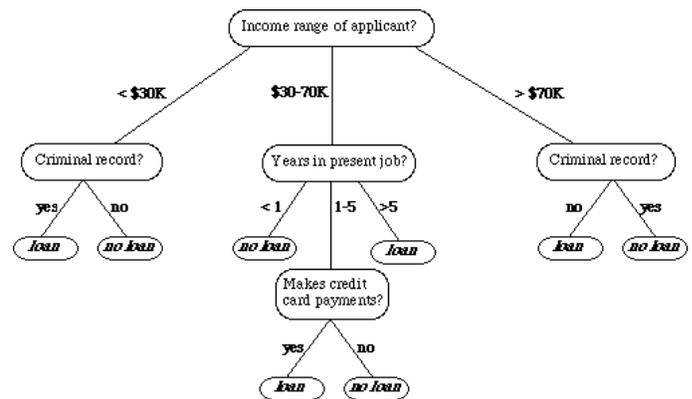

**Figure 4. A visualization of a Decision Tree algorithm evaluating whether or not to approve a loan for an applicant**

1) Decision Trees

Decision Trees are widely utilized within the Machine Learning field as they are straightforward in nature: they simply pose a series of carefully crafted questions in attempts to classify the task, similar to how the popular game '*20 Questions*' works. Yet, there may be hundreds of thousands of possible combinations for these trees, therefore we utilize **Hunt's Algorithm** to populate the trees [28].

**Hunt's Algorithm:**

With Hunt's Algorithm [28], a decision tree is grown in a recursive fashion by partitioning the training records into successively purer (i.e. a subset of the original set where all classes hold the same value). Let $D_t$ represent the set of training records which associate with node *t*, and $y = \{ y_1, y_2, ...., y_c \}$ represent the class labels. Therefore, Hunt's algorithm can become a recursive approach towards solving decision trees through the following:

**Step 1:** If all of the records in $D_t$ belong to the same class $y_t$, we address *t* as a leaf node, and label it $y_t$.

**Step 2:** If $D_t$ contains records which belong to more than one class, we create an *attribute test condition* in order to partition the records into smaller subsets. We create a *child node* for each of these outcomes, and therefore distribute all of the records in $D_t$ to the children based off of the outcomes. We then apply the algorithm recursively to the child node.

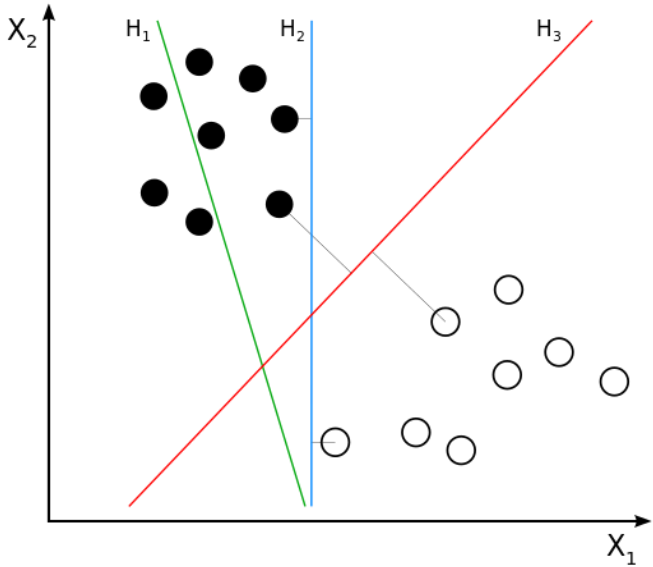

**Figure 5.** Visualization of a Support Vector Machine: if the white and black dots represent the classes, $H_1$ does not separate the classes, $H_2$ minimally separates the classes, and $H_3$ maximally separates the classes.

2) Support Vector Machine Classifiers

A Support Vector Machine (SVM) constructs a hyperplane, or a set thereof within a high-dimensional space, which can be utilized for classification [29]. We use a **Linear SVM**, which simply means we utilize a straight line to differentiate the white dots from the black ones. Algorithmically:

Provided a training dataset of *n* points of form ($X_1$, $Y_1$), ....., ($X_n$, $Y_n$), where $Y_i$ is either 1 or -1, indicating each possible class of which the point $X_i$ may belong. Each $X_i$ is a *p*-dimensional real vector, where we desire to determine the "maximum-margin hyperplane" which divides the group of points $X_i$ for which $Y_i = 1$ from the points for which $Y_i = -1$, such that the distance between the hyperplane and the nearest point $X_i$ from either group is maximized.

We define our hyperplane as a set of points $\vec{X}$ satisfying $\vec{W} \cdot \vec{X} - B = 0$, where $\vec{W}$ is the normal vector to the hyperplane. The parameter $\frac{b}{\|\vec{w}\|}$ determines the offset of the hyperplane from the origin along normal vector $\vec{W}$.

3) Logistic Regression

Developed by Cox *et al.* in 1958, Logistic Regression is a binary logistic model used to estimate the probability of a binary response based one or more predictors (in our case, features) [30]. While it may technically not qualify as a classification method, it represents a discrete choice model, and we therefore use it as such.

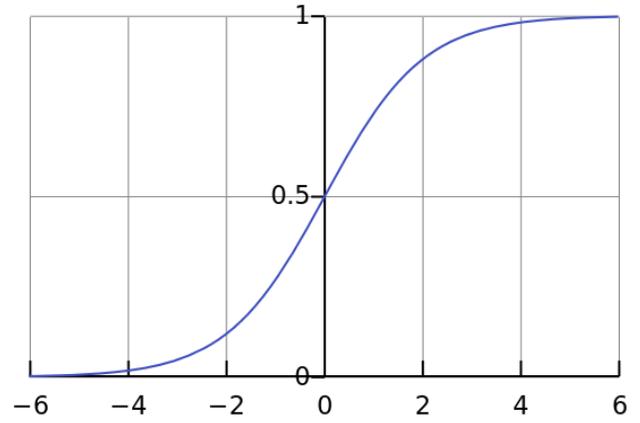

**Figure 6.** Our standard Logistic Function σ(t); the steeper the curve, the more difficult a diagnosis of depressed can be. Therefore, we aim to modify this curve to optimize for accuracy in diagnosis

We define the relationship between our binary dependent variable and our features through equation (1).

$$F(x) = \frac{1}{1 + e^{-(\beta_0 + \beta_1 x)}} \quad (1)$$

In (1), $\beta_0 + \beta_1 x$ represents the parameters of best fit for the success case, hence 'depression'. Therefore, F(x) represents the probability of the dependent variable *t* equaling the depressed case, and inherits a non-depressed bias (i.e. all text is of the control case, unless significant data has been provided to prove otherwise).

4) Naïve Bayes

A Naïve Bayes classifier is one of the simplest available within the Machine Learning field, yet is still competitive with Support Vector Machines, and likes thereof [31]. Based off of the popular Bayes' theorem from statistics, it relies upon an underlying assumption that each feature is *independent* of another, thus vastly simplifying the computational space. For example, a fruit may be classified as an apple if it is red, round, and roughly 10 centimeters in diameter. Under the independence assumption of the Naïve Bayes algorithm, these features would be independent of each other, regardless of any possible correlation between size, shape, and color of a fruit [31].

We apply a Multinomial approach to the Naïve Bayes algorithm. Equations (2) and (3) detail the Naïve Bayes algorithm, while equation (4) details our multinomial approach to Bayes' theorem.

$$p(C_k|\mathbf{x}) = \frac{p(C_k)\ p(\mathbf{x}|C_k)}{p(\mathbf{x})} \quad (2)$$

$$p(C_k|x_1,\ldots,x_n) = \frac{1}{Z} p(C_k) \prod_{i=1}^{n} p(x_i|C_k) \quad (3)$$

| Classification Algorithm | Precision | Recall | F1-Score | Accuracy | Samples |
|---|---|---|---|---|---|
| *Decision Trees* | 0.67 | 0.68 | 0.75 | 0.67 | 332421 |
| *Linear Support Vector Classifier* | 0.83 | ***0.83*** | 0.83 | 0.82 | 332421 |
| *Naïve Bayes w/ 2-grams* | 0.82 | 0.82 | 0.82 | 0.82 | 332421 |
| *Logistic Regression* | ***0.86*** | 0.82 | ***0.84*** | 0.82 | 332421 |
| *Naïve Bayes w / 1-gram* | 0.81 | 0.82 | 0.81 | ***0.86*** | 332421 |
| *Ridge Classifier* | 0.81 | 0.79 | 0.78 | 0.79 | 332421 |

**Table 1: Our generated classification report from various models; bolded text demonstrates the winner in that category.**

$$p(\mathbf{x}|C_k) = \frac{(\sum_i x_i)!}{\prod_i x_i!} \prod_i p_{ki}^{x_i} \quad (4)$$

## V. RESULTS

In this section, we investigate and discuss the degree of accuracy to which the presence of active depression within a body of text may be ascertained from the features extracted from our user's linguistic history. Classifiers were constructed by Machine Learning as detailed in Section 5.3 for estimating the presence of active depression, and we utilized a 6-fold cross-validation to verify our results. We utilize precision, recall, F-measures, and accuracy of the estimation as indices to evaluate depression accuracy, as detailed within Section 4. Lastly, we develop and examine Receiver Operating Characteristic curves to provide an illustration as to the performance of a binary classifier system over various discrimination thresholds.

Table 3 shows the accuracy as to which our constructed classifiers were able to discern the class of a small body of text. The classification accuracies are the average values given by 6-fold cross-validation, and our input was the 846,496 dimensional feature space provided by the Bag of Words approach we used to vectorize the tweet. In implementation, we used employed a CountVectorizer with default settings from Scikit-Learn developed by Pedregosa et al [32]. Table 3 will depict the single scenario in which we analyzed the use of bigrams (ie. frequencies of word couplings, as opposed to singular words) in our classifiers, albeit to no success. As Twitter has implemented a 140-character limit upon their status updates, we found that unigrams were able to capture a significant amount of data present; additionally, a negligible advantage in classification accuracy did simply not outweigh additional computational resources required for a bigram-based approach [33].

Furthermore, Table 3 demonstrates that the presence of active depression within textual bodies may be most accurately estimated with 86% accuracy for a unigram-based Naïve Bayes approach. As Section 5.3.4 detailed, a multinomial approach to Bayes' theorem brought a simple Naïve Bayes classification algorithm up to the likes of Support Vector Machines (SVMs). Nevertheless, while our Naïve Bayes may have produced the best accuracy, it fell behind other models in respect to precision, recall, and F-measure (scoring a 0.81, 0.82, and 0.81 respectively).

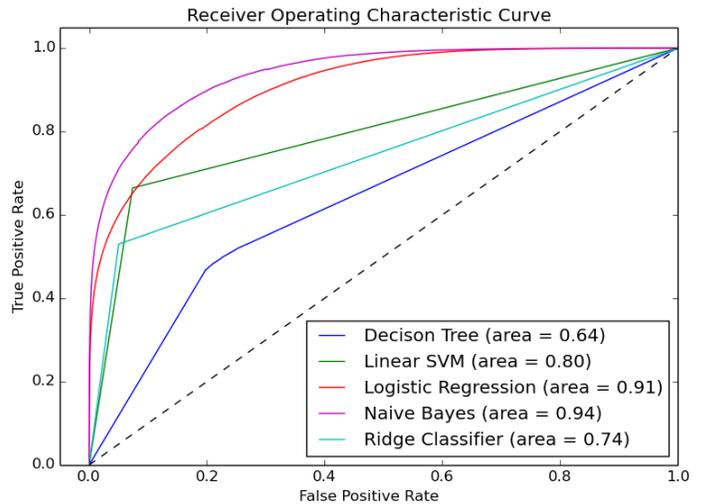

**Figure 7 shows Receiver Operating Characteristic (ROC) curves for our various training models.**

In particular, our Naïve Bayes classifier fell short to Logistic Regression in our classification task. While our Naïve Bayes approach may have attained a precision and F-measure score of 0.81, a Logistic Regression model scored a precision score of 0.86, and outperformed any other classifier with an F1-score of 0.84. A Linear SVM attained the highest recall score (0.83) out of all, but fell behind in precision (0.83) and accuracy (0.82)

$$A = \int_{\infty}^{-\infty} \mathrm{TPR}(T) \mathrm{FPR}'(T)\, dT = \\ \int_{-\infty}^{\infty} \int_{-\infty}^{\infty} I(T' > T) f_1(T') f_0(T)\, dT'\, dT = P(X_1 > X_0) \quad (5)$$

In order to quantitatively measure the performance of a classifier's ROC curve, we investigate its Area under Curve (AUC) metric. As an integral is the de-facto method to easily measure an area under the curve for a function, we develop equation (5) to evaluate each classifier's performance [34]. We found that a Naïve Bayes approach scored the highest out of all our classifiers with a ROC AUC score of 0.94 (Fig. 7). Logistic Regression scored second, with a 0.91, trailing behind were Linear SVMs (0.80), Ridge Classifiers (0.74), and a Decision Tree (0.64). An ROC AUC score of 0.50 is essentially guessing, and an ROC AUC score of 1 is considered perfect [34].

At this point, in order to select a preferred model for further research and potential industrial applications, we must ascertain as to what we desire from our classifier. We defined our measures for an ideal classifier to detect depression as the following:

(1) If prioritization is necessary, we believe in an overbearing approach over an underbearing one, as our research deals with the potential identification of depression rather than treatment. We aim for a control-biased model, and therefore prioritize recall over precision. In lay terms, we emphasize the ability to identifying most depressed individuals at the risk of identifying a few false positives.

(2) We prioritize accuracy over an F1-score: a model which identifies depression well is more important than one which becomes unreliable through a myriad of false positives.

(2.1) As a subset of (2), we evaluate ROC curves alongside accuracy.

(3) Computational resources and time are to be considered, especially if coupled with a decentralized application.

If (1) is to be heeded, a Linear Support Vector Machine would be preferred over all other applications. Yet, (2) prioritizes accuracy over an F-measure, thus reinforcing a Naïve Bayes approach to our classification task. A Naïve Bayes approach attained an average of an 86% accuracy, and consistently performed four points ahead of any other classifier. (2.1) solidifies our choice: a multinomial approach towards a Naïve Bayes classifier is to be explored for further research, as well as potential industrial applications.

## VI. CONCLUSIONS AND FUTURE WORK

We have demonstrated the potential of using twitter as a tool for measuring and predicting major depressive disorder in individuals. First, we compiled a dataset in conjunction with Johns Hopkins University from public self-professions about depression. Next, we proposed a Bag of Words approach towards quantifying this dataset, and created an 846,496 dimensional feature space as our input vector. Finally, we leveraged these distinguishing attributes to build, compare, and contrast several statistical classifiers which may predict the likelihood of depression within an individual.

Our aim was to establish a method by which recognition of depression through analysis of large-scale records of user's linguistic history in social media may be possible, and we yielded promising results with an 86% classification accuracy. The following specific results were obtained: a multinomial approach to the Naïve Bayes' algorithm yielded an A-grade ROC AUC score of 0.94, a precision score of 0.82, and an 86% accuracy; a Bag of Words approach was determined to be a useful feature set, and we determined bigrams to present no significant advantage over a unigram-based approach.

Among future directions, we hope to understand how spatiotemporal behavior may lead to the development of Major Depressive Disorder. The ability to estimate, extrapolate, and interpret daily variations in depression may prove itself as a useful tool for identifying depression prior to mild onset, and therefore expand its potential to save lives. Determining techniques that may be used in a medical context to identify clinical depression from the behavior of social media users' is an important task to benefit the populous.

## VII. ACKNOWLEDGEMENTS

This research paper was made possible through the help and support from everyone, including parents, teachers, family, friends, and in essence: all sentient beings. However, please allow me to especially dedicate my acknowledgement of gratitude towards the following significant advisors and contributors.

First and foremost, I would like to thank Mr. Mike Horn for his support and encouragement. He has pushed me along throughout this process, supported my various undertakings, and has fully revolutionized my senior year in high school. From the bottom of my heart, thank you Mr. Horn.

Secondly, I would like to thank Mr. Glen Coppersmith of Johns Hopkins University, as well as Mr. Will Jack of the Massachusetts Institute of Technology. Mr. Coppersmith has shared with me his dataset, thus making the entire paper possible, as well as provided invaluable insight, guidance, and support in a time of dire need. Mr. Jack has provided a tender hand during late hours, and helped me understand various Machine Learning principles which otherwise were alien to me. I began this paper quite clueless about the realm of Machine Learning technologies, but as a result of Mr. Coppersmith's and Mr. Jack's guidance, have not only been able to survive but also thrive.

Finally, I would like to thank Dr. Sandip Sen at the University of Tulsa. He took my young, frankly naïve mind in, and fostered a level of curiosity which brought me here. Dr. Sen opened my mind to the realm of computational sciences, and I would not be where I am today without him.

I sincerely thank my mentors, parents, family, and friends who have advised and pushed me forward throughout this endeavor. The product of this research paper would not be possible without all of them.